\documentclass{article}
\usepackage{xcolor}
\usepackage{booktabs,multirow,array}
\usepackage[T1]{fontenc}
\usepackage{amsmath}
\usepackage{amssymb}
\usepackage{bm}
\usepackage{physics}
\usepackage{subcaption}
\usepackage{mathtools}
\usepackage{etoolbox}
\usepackage{hyperref}
\usepackage{url}
\usepackage{xspace}
\usepackage{silence}
\newcommand{\ours}{FreNBRDF\xspace}
\newcommand{\intd}{\mathrm{d}}

\AtEndPreamble{
    \usepackage[capitalize]{cleveref}
    \crefname{section}{Sec.}{Secs.}
    \Crefname{section}{Section}{Sections}
    \Crefname{table}{Table}{Tables}
    \crefname{table}{Tab.}{Tabs.}
    \Crefname{algorithm}{Algorithm}{Algorithms}
    \crefname{algorithm}{Alg.}{Algs.}
    \Crefname{figure}{Figure}{Figures}
    \crefname{figure}{Fig.}{Figs.}
}

\RequirePackage{enumitem}
\setlist[itemize]{noitemsep,leftmargin=*,topsep=0em}
\setlist[enumerate]{noitemsep,leftmargin=*,topsep=0em}

\makeatletter
\DeclareRobustCommand\onedot{\futurelet\@let@token\@onedot}
\def\@onedot{\ifx\@let@token.\else.\null\fi\xspace}

\def\eg{\emph{e.g}\onedot}

\def\etal{\emph{et al}\onedot}
\makeatother

\usepackage{algorithm, algpseudocode}
\usepackage{pdfpages}

\usepackage{longtable,booktabs,array}
\usepackage{fancyvrb}

\DefineVerbatimEnvironment{Highlighting}{Verbatim}{commandchars=\\\{\}}

\usepackage{amsmath,graphicx,mlspconf}
\usepackage{xcolor}

\toappear{2025 IEEE International Workshop on Machine Learning for Signal Processing, Aug.\ 31-- Sep.\ 3, 2025, Istanbul, Turkey}

\title{\ours: A Frequency-Rectified Neural Material Representation}
\name{Chenliang Zhou$^{\dagger}$\thanks{$^{\dagger}$: Equal contribution.}, Zheyuan Hu$^{\dagger}$\footnotemark[1], Cengiz Oztireli}
\address{Department of Computer Science and Technology\\University of Cambridge}

\usepackage{graphicx}
\usepackage{etoolbox}
\newcommand{\insertfig}{\includegraphics[width=0.9\linewidth]{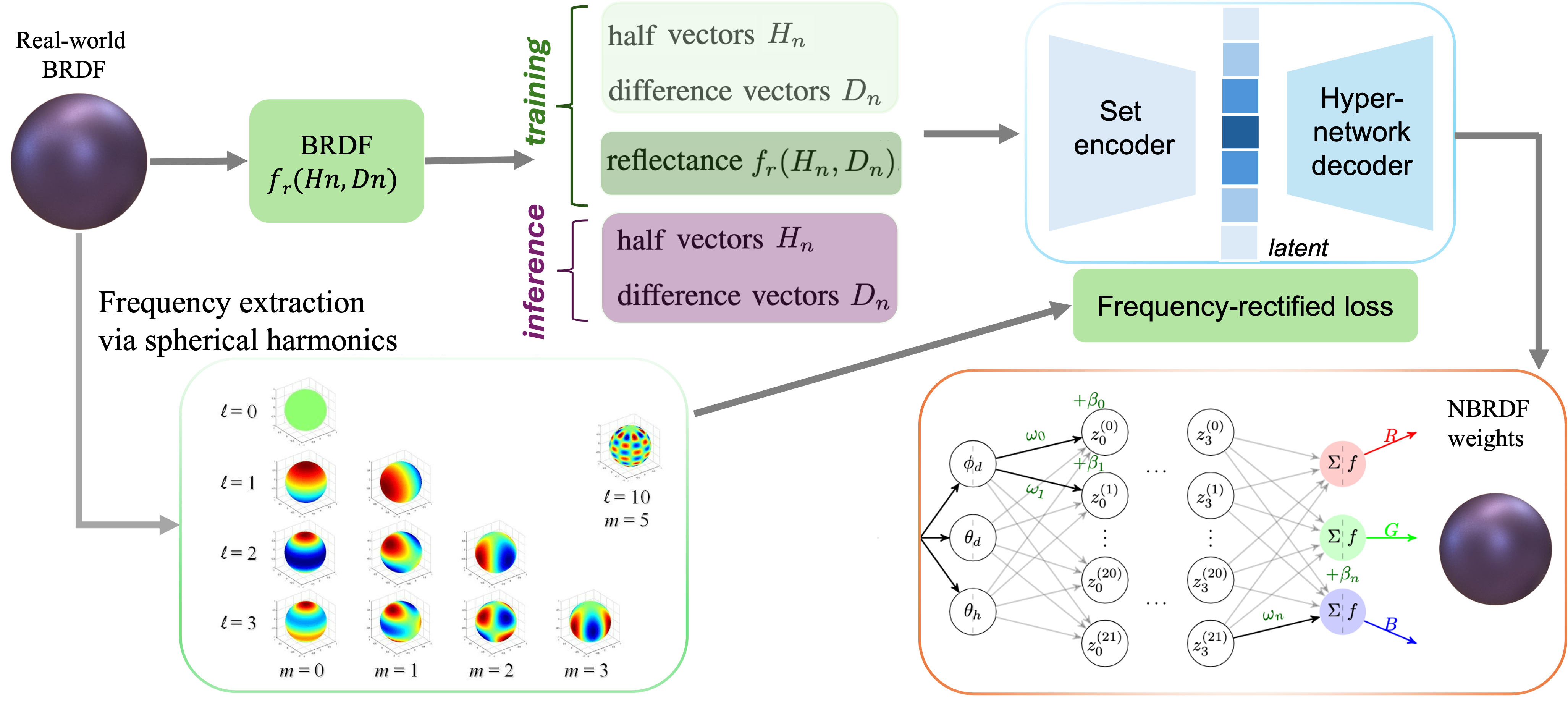}\captionof{figure}{Overview of \ours architecture.}\label{fig:reconstruct-hypernetwork}}% define the image

\makeatletter
\apptocmd{\@maketitle}{\centering\insertfig\medskip}{}{}% insert the figure after authors
\makeatother
\begin{document}
%\ninept

\maketitle

\begin{abstract}
Accurate material modeling is crucial for achieving photorealistic rendering, bridging the gap between computer-generated imagery and real-world photographs. While traditional approaches rely on tabulated BRDF data, recent work has shifted towards implicit neural representations, which offer compact and flexible frameworks for a range of tasks. However, their behavior in the frequency domain remains poorly understood. To address this, we introduce \emph{\ours}, a frequency-rectified neural material representation. 
% We present a comprehensive frequency analysis of the material reconstruction pipeline, providing deeper insights into the spectral properties of neural BRDF representations. Furthermore, we introduce
By leveraging spherical harmonics, we integrate frequency-domain considerations into neural BRDF modeling. We propose a novel \emph{frequency-rectified loss}, derived from a frequency analysis of neural materials, and incorporate it into a generalizable and adaptive reconstruction and editing pipeline. This framework enhances fidelity, adaptability, and efficiency. Extensive experiments demonstrate that \ours improves the accuracy and robustness of material appearance reconstruction and editing compared to state-of-the-art baselines, enabling more structured and interpretable downstream tasks and applications.
\end{abstract}
\begin{keywords}
Appearance modeling, frequency-aware methods, material reconstruction, material editing
\end{keywords}

\section{Introduction} \label{chap:intro}
Material properties play a crucial role in visual computing, serving as a fundamental component in applications such as rendering, augmented reality, and scene understanding \cite{zhou2024neumadiffneuralmaterialsynthesis}. This is usually realized through the modeling and reconstruction of the \emph{bidirectional reflectance distribution functions (BRDFs)}~\cite{brdf_evaluation_techniques21}, which describe the relationship between the incident and outgoing radiances for a specific material. By capturing the complex interactions of light with different surfaces, material representations enhance the realism and fidelity of synthesized images, improving the overall visual quality in various computational applications.

% BRDF describes the interaction of light with a point on a geometric surface, which is widely used in Physically Based Rendering. Analytical ~\cite{phong, lafortune, ward, cook-torrance, ggx} and data-driven measured ~\cite{Matusik2003datadriven} BRDFs are two main categories. Each model tradeoffs material fidelity, performance (i.e. time and memory) costs  and ease of understanding and deployment. 

Data-driven BRDF models~\cite{Matusik2003datadriven}, which map spatial coordinates to reflected light distributions, have demonstrated high fidelity in capturing complex material appearances. However, their inherently large data size poses significant challenges for downstream tasks such as material reconstruction, editing, storage, and real-time rendering. Efficiently representing and reconstructing BRDF data still remains a critical issue.

Implicit neural BRDF models (\eg, \emph{NBRDFs}~\cite{sztrajman2021nbrdf, zhou2024neumadiffneuralmaterialsynthesis, zhou2024physically}) offer a promising solution to these challenges by incorporating neural networks into the BRDF modeling pipeline, providing a more scalable and adaptive solution for material representation. Based on NBRDF, recent research has explored leveraging latent space representations through an encoder-decoder architecture~\cite{gokbudak2023hypernetworks}. This approach seeks to reconstruct BRDFs from a compressed latent space, reducing memory requirements while maintaining high reconstruction fidelity.

However, the behavior of material representations in the frequency domain remains insufficiently explored. To address this gap, we propose \emph{\ours}, a frequency-rectified neural material representation. By leveraging spherical harmonics, we integrate frequency rectification~\cite{zhou2023frepolad} into the NBRDF framework and introduce a novel \emph{frequency-rectified loss} to achieve a more structured and accurate representation of material appearance. The primary contributions of this work are:
\begin{itemize}
    \item A method for extracting frequency information from neural materials using spherical harmonics; and  
    \item A generalizable and adaptive material reconstruction and editing framework based on an autoencoder architecture, incorporating frequency rectification and achieving state-of-the-art performance.
\end{itemize}

\section{Related Work}
\textbf{Material acquisition}\hspace{1ex} Gonio-photometers~\cite{PAB2018} are widely used for measuring retro-reflection, enabling precise characterization of material appearance. Using this technique, RGL~\cite{Dupuy2018Adaptive} captured 62 materials, including anisotropic ones. In contrast, image-based measurement devices~\cite{Marschner2000} were used to construct the MERL dataset~\cite{Matusik2003datadriven}, which consists of 100 isotropic materials. Similarly, digital consumer cameras have been employed for texture acquisition~\cite{schwartz-2014-setups}, forming the \emph{bidirectional texture function database Bonn (BTFDBB)} using 151 such cameras mounted on a structured rack.  

\smallskip \noindent \textbf{Material modeling}\hspace{1ex} \emph{Bidirectional reflectance distribution functions (BRDFs)}~\cite{brdf_evaluation_techniques21} are widely adopted for material modeling. Early BRDF models~\cite{phong, lafortune, ward} introduced analytical expressions derived from empirical observations. Other models~\cite{cook-torrance, ggx, disney} incorporate physical scattering principles while maintaining artist-friendly parameterization. These models offer a balance between realism and efficiency, as their limited set of tunable parameters can be fitted to measured material data~\cite{disney}, making them computationally efficient in terms of both time and memory. However, due to inherent model approximations, many analytical BRDFs struggle to capture complex reflectance behaviors, particularly for highly detailed or anisotropic materials~\cite{sztrajman2021nbrdf}.

\smallskip \noindent \textbf{Neural BRDFs}\hspace{1ex} Data-driven material representations (\eg, MERL~\cite{Matusik2003datadriven}) produce highly realistic renderings compared to analytical models. However, their large data size presents a significant challenge for real-world applications.
% \footnote{For example, each BRDF in the MERL dataset is stored using $90 \times 90 \times 180 \times 3$ floating-point numbers.}.
To address this, compression techniques such as dimensionality reduction~\cite{NielsenPCA2015} or neural network-based compression~\cite{sztrajman2021nbrdf, gokbudak2023hypernetworks} are commonly used. Among these, \emph{neural BRDFs (NBRDFs)}~\cite{sztrajman2021nbrdf} offer a compelling balance of high fidelity and low memory demand, making them an efficient latent representation for various tasks. Built upon this, recent research has introduced a set encoder hypernetwork~\cite{gokbudak2023hypernetworks} that enables generalizable and adaptive BRDF reconstruction from sparse samples. This approach is designed to handle unseen data while accommodating varying sample sizes, making it a flexible and scalable solution. Building on this work, we further explore neural BRDF representations within the frequency domain.
% and enhance it through targeted evaluations and improvements.  

\smallskip \noindent \textbf{Frequency rectification}\hspace{1ex} FrePolad~\cite{zhou2023frepolad} utilizes spherical harmonics~\cite{bonev2023spherical} to extract frequency information from point clouds, improving the quality and diversity of point cloud generation while maintaining high computational efficiency. Inspired by this approach, we integrate frequency information into the NBRDF representation, aiming to enhance accuracy in NBRDF modeling, reconstruction, and editing.
\section{Method}
\label{chap:method}

\subsection{NBRDF autoencoder pipeline}
\label{sec:NBRDF autoencoder pipeline}
Under Rusinkiewicz reparameterization of BRDFs~\cite{coordinate98}, we represent the incoming and outgoing directions using the half $H$ and difference $D$ vectors or their spherical coordinates \(\theta_H, \phi_H\) and the difference vector \(\theta_D, \phi_D\), respectively.
% Data-driven BRDFs \(f_r\) maps \((\theta_H, \theta_D, \phi_D)\) to reflectance values in \(\mathbb{R}^3\). These functions are usually measured with high dynamic range (HDR) terms.
Note that for isotropic materials, the BRDFs are independent of \(\phi_D\). Therefore, a BRDF can be expressed as either $f_r(H, D)$ or $f_r(\theta_H, \theta_D, \phi_H)$.

Inspired by prior work~\cite{sztrajman2021nbrdf, gokbudak2023hypernetworks, zhou2024physically}, we first represent BRDFs via lightweight neural fields called \emph{NBRDFs}~\cite{sztrajman2021nbrdf} that map the half \(H_n\) and difference vectors \(D_n\) to reflectance values \(f_r(H_n, D_n)\). To facilitate NBRDF reconstruction and editing, our autoencoder-based pipeline trains a pair of encoder and decoder to capture a lower dimensional space of the weights of NBRDFs. During inference, the model accepts queries in terms of $(H,D)$ coordinate and returns the reflectance values at that coordinate. Moreover, the set encoder provides low-dimensional embeddings of neural materials, enabling material editing through interpolation. By linearly interpolating between the embeddings of different materials, we can reconstruct a range of novel appearances. We illustrate this autoencoder architecture in \cref{fig:reconstruct-hypernetwork}.

For the network, we adopt a set encoder~\cite{zaheer2017deepset} with permutation invariance and flexibility of input size. It
takes an arbitrary set of samples as input, which is the concatenation of BRDF values and coordinates, containing four fully connected layers with two hidden layers of 128 hidden neurons and ReLU activation function. The reconstruction loss $\mathcal{L}_{\text{rec}}(f_r, f_r')$ between two NBRDFs is defined to be the sum of the L1 loss between samples of the two underlying BRDFs and the two regularization terms for NBRDF weights \(w\) and latent embeddings \(z\):
\begin{align}
\label{eq:baseline-loss}
\nonumber \mathcal{L}_{\text{rec}}(f_r, f_r') \coloneqq  & \frac{1}{|P|} \sum_{i=1}^{|P|} \left|\left( f_{r}(H_i, D_i) - f_{r}'(H_i, D_i) \right) \cos \theta_{i} \right| \\ &+ \lambda_1 \sum_{i=1}^{W} w_i^2 + \lambda_2 \sum_{j=1}^{Z} z_j^2,
\end{align}
where \(P\) is the predefined set of indices for samples, \(\theta_i\) is the angle between the incoming light direction and the normal of the surface, and $\lambda_1$ and $\lambda_2$ are regularization coefficients.

Following previous work~\cite{NielsenPCA2015}, we also adopt the pre-processing technique of log-relative mapping for BRDFs:
\begin{equation}
f_r \mapsto \ln \left( \frac{f_r   + \epsilon}{f_{\text{ref}} \, + \epsilon} \right)
\end{equation}
where the reference value \(f_{\text{ref}}\) is the median BRDF value and \(\epsilon\) is a small constant to avoid division or logarithm of zero. 
% Note that we also tried taking the average value for \(f_{\text{ref}}\) instead and the result is evaluated as well.

\subsection{Frequency Recitification}
\label{algorithm-improvements}
We observe that the naive NBRDF autoencoder pipeline described above lacks the frequency information between the two NBRDFs, which might help the learning of NBRDF weight distribution~\cite{zhou2023frepolad}. To extract frequency information from BRDFs, we apply spherical harmonic transformation~\cite{bonev2023spherical, zhou2023frepolad}. Note that the spherical harmonics require continuous functions on the unit sphere $S^2$. Our first problem is that the BRDF $f_r(\theta_H, \theta_D, \phi_H)$ is not a function on $S^2$. We address this by fixing the third argument $\phi_H$ to some value $\alpha$, and the function is thus defined on $S^2$:
\begin{equation}
    f_r(\theta, \phi) \coloneqq f_r(\theta_H, \theta_D, \phi_H=\alpha).
\end{equation}
We can then apply the technique multiple times for different values of $\phi_H$ and average the result.

Another problem is that the given MERL dataset~\cite{Matusik2003datadriven} only provides BRDF values at discrete sample positions $P$, but we need a continuous function. To overcome this, we obtain the BRDF values at an arbitrary position $(\theta, \phi)$ by interpolating the values at its $k$ nearest neighbors $P_k(\theta, \phi)$ in the sample positions $P$ with weights $w_i$:
\begin{equation}
f_r(\theta, \phi) \coloneqq
\sum_{i \in P_k(\theta, \phi)} w_i f_r(\theta_i, \phi_i).
\end{equation}
% \begin{equation}
% g_{Mat}(\theta_D, \phi_D) = 
% \begin{cases} 
% f_r, & \text{if } (\theta_D, \phi_D, f_r) \in Mat \\
% \sum_{(\theta_i, \phi_i) \in kNN_{(\theta_D, \phi_D)}} \omega_i(\theta_i, \phi_i) \cdot f_{r_i}, & \text{otherwise}
% \end{cases}
% \end{equation}
The $k$ nearest neighbors $(\theta_i, \phi_i)$ of $(\theta, \phi)$ are determined through their distances $d_i$ to the points $(\theta, \phi)$:
\begin{equation}
d_i = \sqrt{2 - 2 \left[ \sin \theta \sin \theta_i \cos(\phi-\phi_i) + \cos \theta \cos \theta_i  \right]}.
\end{equation}
For the choice of weights $w_i$, since closer points should have larger weights, $\{w_i\}_i$ should be a decreasing sequence on $d_i$. Consequently, a suitable candidate is the normalized Gaussian function with standard deviation $\sigma$:
\begin{align}
    w_i' &\coloneqq e^{-\frac{di^2}{2 \sigma^2}}; \\
    w_i &\coloneqq \frac{w_{i}'}{\sum_j w_{j}'}.
\end{align}

After we obtain BRDFs expressed as continuous functions on the unit sphere $S^2$, we can further leverage spherical harmonics to express them as a linear combination of orthonormal base functions
\(G: S^2 \rightarrow \mathbb{C}\) of degrees
\(l\) and orders \(m\):
\begin{align}
f_r(\theta, \varphi) &= \sum_{l=0}^\infty \sum_{m=-l}^l c_{l,m} G_{l,m}(\theta, \varphi); \\
c_{l,m} &= \int_{0}^{2\pi} \int_{0}^{\pi} f_r(\theta, \phi) \overline{G _{l,m}(\theta, \phi)}\sin\theta \intd \theta \intd \phi.
\end{align}
Note that BRDFs have three channels (RGB). Therefore, we apply the above method to each channel separately, producing the frequency coefficients $c_{l,m} \in \mathbb{R}^3$. The key insight here is that these frequency coefficients now contain the extracted frequency information at each degree $l$ and order $m$. Therefore, we can define a \emph{frequency-rectified loss} on BRDFs based on the mean squared error of frequency coefficients
\begin{equation}
\label{eq:L-fre}
    \mathcal{L}_\text{fre}(f_r, f_r') \coloneqq \frac{1}{|P|}\sum_{i=1}^{|P|} \left\|c_{l,m} - c_{l,m}' \right\|^2
\end{equation}
and consequently, we can incorporate this loss into the reconstruction loss \cref{eq:baseline-loss}:
\begin{equation}
\label{eq:L-total}
    \mathcal{L}(f_r, f_r') \coloneqq \mathcal{L}_\text{rec}(f_r, f_r') + \eta \mathcal{L}_\text{fre}(f_r, f_r'),
\end{equation}
where $\eta \in \mathbb{R}$ is the hyperparameter controlling the strength of frequency rectification and $c_{l,m}$ and $c_{l,m}'$ are the frequency coefficients for $f_r$ and $f_r'$, respectively. With the help of frequency information, our pipeline is able to better learn the NBRDF weight distribution, enabling better reconstruction and editing tasks as demonstrated in \cref{chap:experiments}.

% An immediate attempt to incorporate this frequency information involves
% reshaping the BRDF, such that the three-dimensional positional vectors are mapped into the spherical harmonic transformation two-dimensional input space, as the frequency function to be extracted. 
% % The relevant details are left in the \cref{implementation-details}.
% We then replace the
% \(\mathcal{L}_{\text{rec}}\) with only the frequency loss 
% \(\text{MSE}(c^{pred}_{l,m} , c^{true}_{l,m})\).

% After extracting the frequency information, we apply it as guidance added to the original reconstruction loss with scale \(\eta = 10\), forming
% \begin{equation}
% \mathcal{L}_{fre} = \mathcal{L} + \eta \cdot \text{MSE}(c^{pred}_{l,m} , c^{true}_{l,m}).
% \label{eq:L-freq}
% \end{equation}

% The \texttt{Dataset} class, derived from the \texttt{Pytorch}, returns the Cartesian positions of half \(H_n\) and difference vectors \(D_n\), along with reflectance measurements \(f_r (H_n, D_n)\) , to be used by the
% training pipeline via the \texttt{Dataloader} class.

\begin{figure*}
    \centering
    \includegraphics[width=1\linewidth]{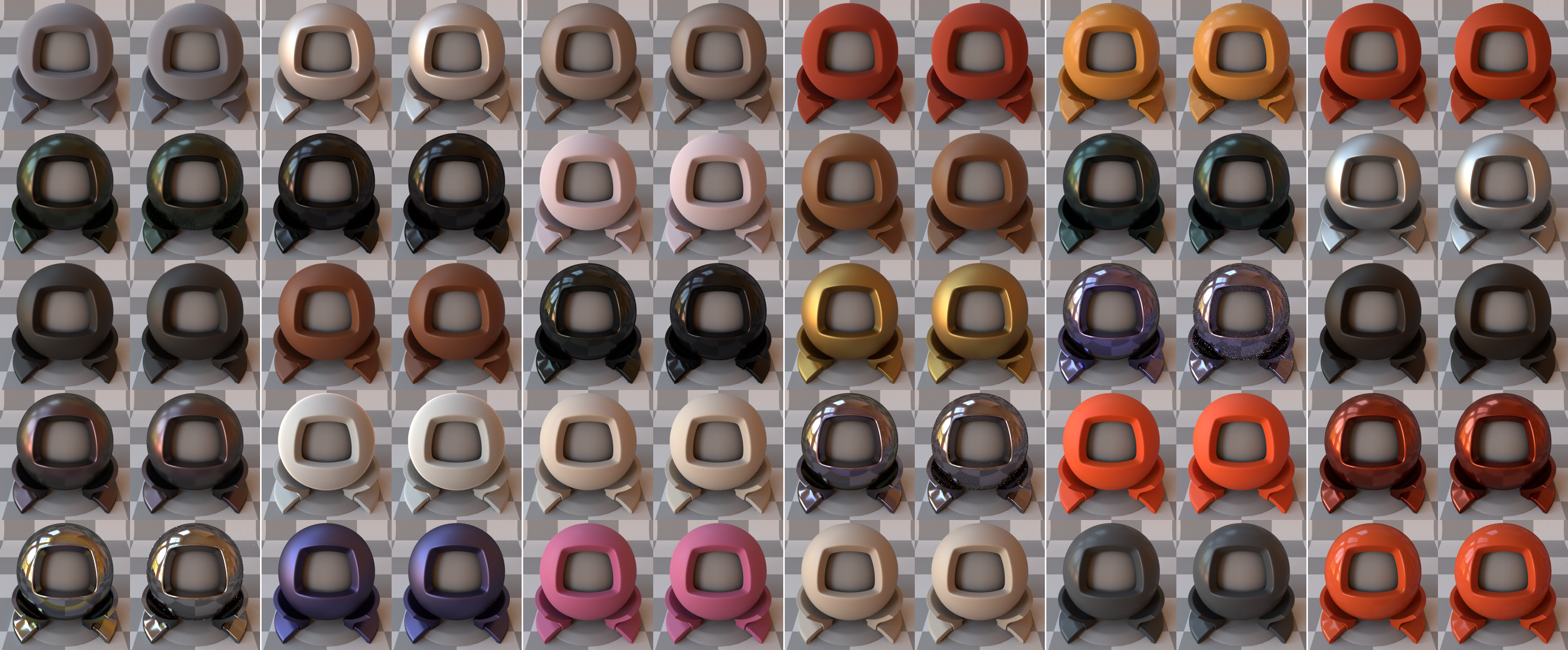}
    \caption{\ours reconstructs 30 MERL materials~\cite{Matusik2003datadriven} with high quality, indicating that \ours effectively learns the material distribution (ground truths to the left and the reconstructed material to the right).}
    \label{fig:reconstruction}
\end{figure*}
\section{Experiments}
\label{chap:experiments}
\subsection{Dataset}\label{datasets}
The MERL dataset~\cite{Matusik2003datadriven} is used in this study for its diversity and data-driven nature, making it suitable for both statistical and neural-network-based methods. It contains 100 measured real-world materials. Each BRDF is represented as a \(90 \times 90 \times 180 \times 3\) floating-point array, mapping uniformly sampled input angles $(\theta_H, \theta_D, \phi_D)$ under Rusinkiewicz reparametrization~~\cite{coordinate98} to reflectance values in \( \mathbb{R}^3 \).

\subsection{Training and testing procedures}
\label{training-and-testing-procedures}
The dataset is divided into training, validation, and testing sets in a 70\%-10\%-20\% split. We use the Adam optimizer~\cite{Kingma2014AdamAM}, a first-order gradient-based method for optimizing stochastic objective functions.

During training, the specified losses are reported at each epoch, and optimal hyperparameters are selected based on performance on the validation set. Model performance is then evaluated on the hold-out test set using a range of quantitative metrics introduced in \cref{sec:matrics}.
% Please refer to \cref{training-and-testing-results} for a detailed training report.

% Hyperparameter tuning by grid search. For example, learning rate iterates from 1e-1, 1e-3 to 1e-5. The chosen parameter, with the best performance, serves as the random search interval for finding the precise value.

% Checkpoints are used for backup and reproducibility. 
% Hyperparameter, training status are saved together with model weights. They provide the foundation for me to compare between models throughout the project and
% resume from history rather than training again from scratch.

\subsection{Metrics}
\label{sec:matrics}
To evaluate the performance of material reconstruction and editing, we report the frequency-rectified loss (\cref{eq:L-fre}) accessing the frequency compliance. We also report the following three rendering-based metrics assessing the perceptual similarity and reconstruction quality between the two images \(I_1, I_2\colon \{1,2,\ldots,w\} \times \{1,2,\ldots,h\} \rightarrow [0,1]^3\), where $w,h$ are the width and height of the images:
\begin{itemize}
\item Root mean squared error (RMSE). RMSE checks if pixel values at the same coordinates match:
\begin{equation}
\mathcal{L}_\text{RMSE}(I_1,I_2) \coloneqq \sqrt{\frac{1}{wh} {\sum_{i=1}^{w} \sum_{j=1}^{h} \left( I_1(i,j)-I_2(i,j) \right)^2} }.
\end{equation}
Note that RMSE depends strongly on the image intensity scaling. RMSE aims for a lower value.

\item Peak signal-to-noise ratio (PSNR). PSNR is the scaled mean squared error (MSE). PSNR measures the image reconstruction quality and aims for higher values. Given the
maximum pixel value \(p\), PSNR is defined as
\begin{equation}
\mathcal{L}_\text{PSNR}(I_1, I_2) \coloneqq  10 \log_{10} \frac{p^2}{\mathcal{L}_\text{RMSE}^2(I_1, I_2)}.
\end{equation}

\item Structural similarity index measure (SSIM)~\cite{SSIM04}. SSIM is a perception-based metric that measures the perceptual similarity between the two images, which aims for higher values for better performance. SSIM is computed from the luminance $l(I_1, I_2)$, contrast $c(I_1, I_2)$, and structure $s(I_1, I_2)$ of the two images
% \begin{align}
%    & l(I_1, I_2) \coloneqq \frac{2 \mu_{I_1} \mu_{I_2} + c_1}{\mu_{I_1}^2 + \mu_{I_2}^2 + c_1};\\
%    & c(I_1, I_2) \coloneqq \frac{2 \sigma_{I_1} \sigma_{I_2} + c_2}{\sigma_{I_1}^2 + \sigma_{I_2}^2 +c_2 };\\
%    & s(I_1, I_2) \coloneqq \frac{\sigma_{I_1 I_2} + c_3}{\sigma_{I_1} \sigma_{I_2} + c_3},
% \end{align}
% where $\mu_I, \sigma_I$, and $\sigma_{I_1I_2}$ are the pixel sample mean, sample standard deviation, and sample covariance of the images:
% \begin{align}
%     & \mu_I \coloneqq  \frac{1}{w \cdot h} {\sum_{i=1}^{w} \sum_{j=1}^{h}  I(i,j) };\\
%     & \sigma_I \coloneqq \sqrt{ \frac{1}{w \cdot h- 1} {\sum_{i=1}^{w} \sum_{j=1}^{h}  \left( I(i,j)-\mu_I \right)^2 }};\\
%     & \sigma_{I_1I_2} \coloneqq \frac{1}{w \cdot h - 1} \sum_{i=1}^{w} \sum_{j=1}^h \left(I_1(i,j)-\mu_{I_1})(I_2(i,j)-\mu_{I_2} \right),
% \end{align}
% and $c_1 \coloneqq (k_1 L)^2$, $c_2 \coloneqq (k_2 L)^2$,   $c_3 \coloneqq \frac{c_2}{2}$ are the variables to stabilize the division with weak denominator where $k_1$ and $k_2$ are coefficients defaulted to 0.01 and 0.03, respectively, and $L$ is the dynamic range of the pixel values, which is typically chosen to be $2^l-1$ where $l$ is the number of bits encoding each pixel. Finally, SSIM is defined as a weighted combination of the above comparative measures
with exponential weights $a,b,c > 0$:
\begin{equation}
\mathcal{L}_\text{SSIM}(I_1, I_2) \coloneqq l^a(I_1, I_2) c^b(I_1, I_2) s^c(I_1, I_2).
\end{equation}

% \begin{align}
%     & c_1 \coloneqq (k_1 L)^2; \\
%     & c_2 \coloneqq (k_2 L)^2; \\
%     & c_3 \coloneqq \frac{c_2}{2}
% \end{align}
\end{itemize}

\subsection{Material reconstruction}
We present 30 reconstructed materials rendered under consistent scene and lighting conditions in \cref{fig:reconstruction}. The results demonstrate that \ours effectively learns the material distribution and produces high-quality, faithful reconstructions. In \cref{tab:reconstruction}, we compare the performance of \ours on the material reconstruction task with two state-of-the-art baselines: the method of Gokbudak \etal~\cite{gokbudak2023hypernetworks} and a naive NBRDF~\cite{sztrajman2021nbrdf} reconstruction pipeline, as described in \cref{sec:NBRDF autoencoder pipeline}, without frequency rectification. From the results, we can see that \ours outperforms both baselines across most metrics evaluating frequency compliance, rendering quality, and visual similarity, confirming the effectiveness of our proposed approach. 
Note that the higher RMSE score for \ours is likely due to its reconstruction loss (\cref{eq:L-total}) being designed to enforce consistency in both the spatial and frequency domains.
\begin{table}
    \centering
    \begin{tabular}{rccc}
        \toprule
        Metrics & \cite{gokbudak2023hypernetworks} & NBRDF~\cite{sztrajman2021nbrdf} & \ours (ours) \\
        \midrule
        % data rows here
        RMSE$\times 10^2$ (↓) & 6.89 & \textbf{6.60} & 6.74 \\
        PSNR (↑) & 26.3 & 29.2 & \textbf{29.9} \\
        SSIM$\times 10$ (↑) & 9.26 & 9.50 & \textbf{9.88} \\
        $\mathcal{L}_\text{fre} \times 10^3$ (↓) & 6.74 & 6.80 & \textbf{0.23} \\
        \bottomrule
    \end{tabular}
    \caption{Quantitative comparison on material reconstruction against state-of-the-art baselines. \ours achieves superior performance, showing its effectiveness.}
    \label{tab:reconstruction}
\end{table}
 
\subsection{Material editing}
\label{quantitative-results}
Our pipeline provides a low-dimensional space of neural materials, enabling material editing by linearly interpolating between embeddings of different materials. We compare different models on this task. The ground truth can be obtained by directly linearly interpolating the MERL materials~\cite{Matusik2003datadriven}.

\begin{figure}
    \centering
    \includegraphics[width=1\linewidth]{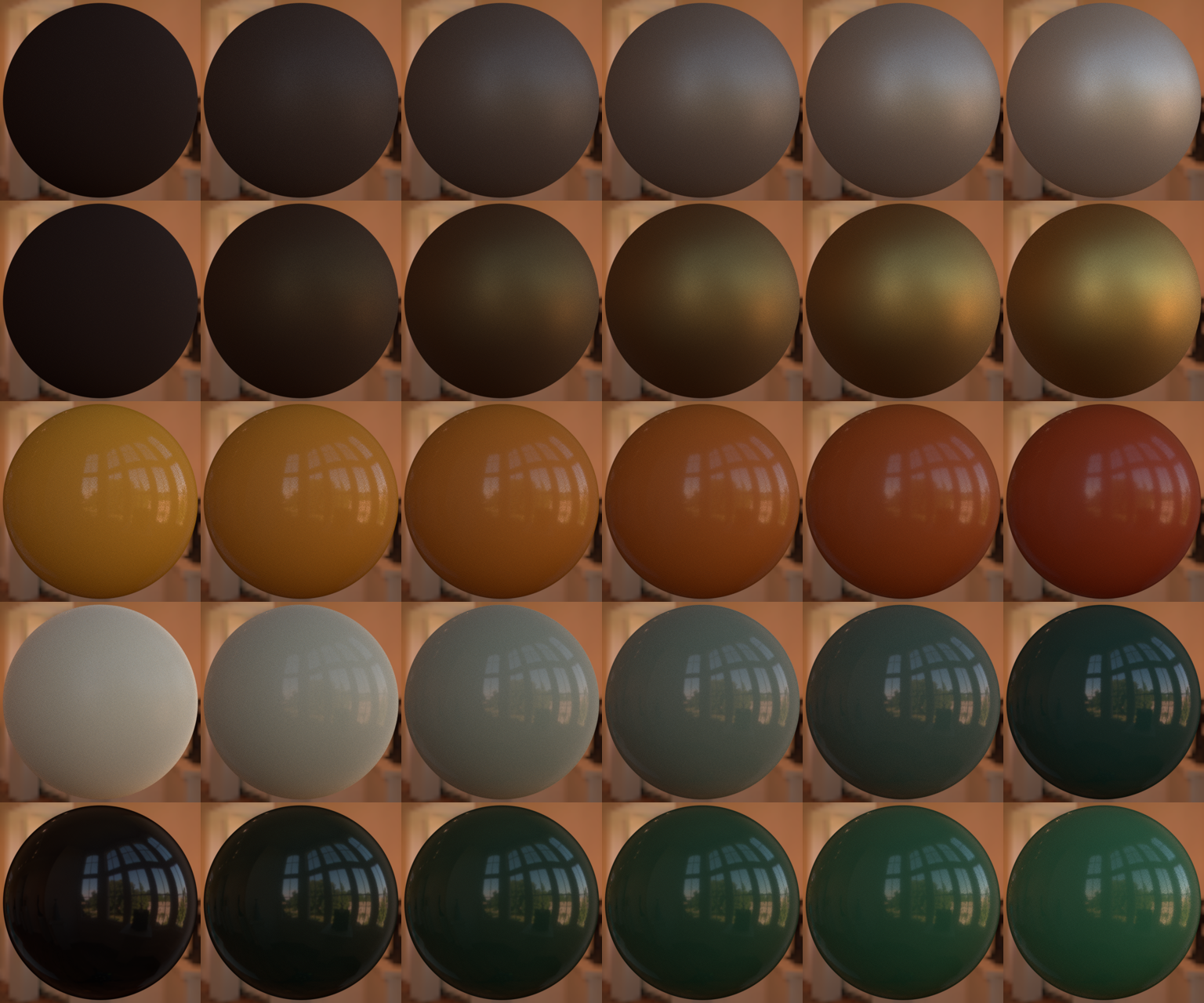}
    \caption{Linearly interpolation between embeddings of five pairs of MERL materials (each row is an interpolation). From left to right, the interpolation weights are 0.0, 0.2, 0.4, 0.6, 0.8, 1.0. The smooth transitions between the two endpoints demonstrate the capability of our \ours as a robust and effective implicit neural representation for materials.}
    \label{fig:interpolation}
\end{figure}

\Cref{fig:interpolation} illustrates interpolations between five pairs of MERL materials where each row represents one interpolation. The smooth transitions between the two endpoints demonstrate the capability of our \ours as a robust and effective implicit neural representation for materials. We also report the relevant metrics in \cref{tab:interpolation}, computed over 2000 randomly interpolated materials. The results show that materials represented by \ours exhibit consistently higher quality compared to the two baselines. Compared to the reconstruction results in \cref{tab:reconstruction}, the interpolated materials produced by the two baselines show degraded performance, while those generated by \ours maintain similar quality. This indicates that \ours effectively captures the underlying distribution of neural materials.
\begin{table}
    \centering
    \begin{tabular}{rccc}
        \toprule
        Metrics & \cite{gokbudak2023hypernetworks} & NBRDF~\cite{sztrajman2021nbrdf} & \ours (ours) \\
        \midrule
        RMSE$\times 10^2$ (↓) & 8.42 & 7.63 & \textbf{6.92} \\
        PSNR (↑) & 22.4 & 27.6 & \textbf{30.1} \\
        SSIM$\times 10$ (↑) & 9.13 & 9.42 & \textbf{9.82} \\
        $\mathcal{L}_\text{fre} \times 10^3$ (↓) & 6.72 & 6.80 & \textbf{0.22} \\
        \bottomrule
    \end{tabular}
    \caption{Quantitative comparison on material editing with two state-of-the-art baselines. Materials represented by \ours exhibit significantly higher quality compared to the two baselines,
    % Compared to the reconstruction results (\cref{tab:reconstruction}), the interpolated materials by the two baselines obtain worse scores whereas those by \ours maintain similar quality,
    indicating that \ours effectively learns the distribution of neural materials.
    }
    \label{tab:interpolation}
\end{table}

\subsection{Further analysis}
\begin{figure}
\centering
\begin{subfigure}{0.3\linewidth}
        \includegraphics[width=\linewidth]{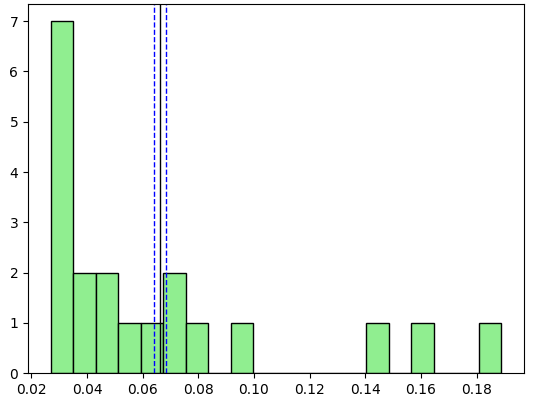}
    \caption{RMSE}
\end{subfigure}
\begin{subfigure}{0.3\linewidth}
        \includegraphics[width=\linewidth]{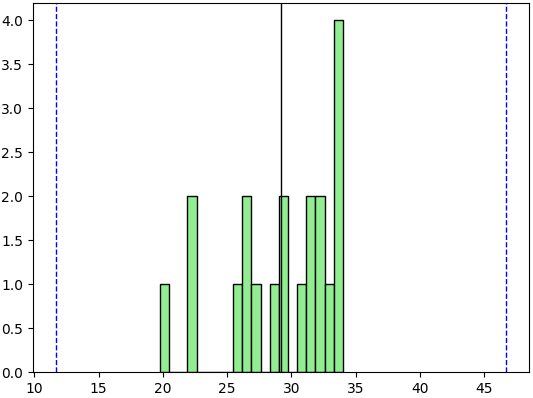}
    \caption{PSNR}
\end{subfigure}
\begin{subfigure}{0.3\linewidth}
        \includegraphics[width=\linewidth]{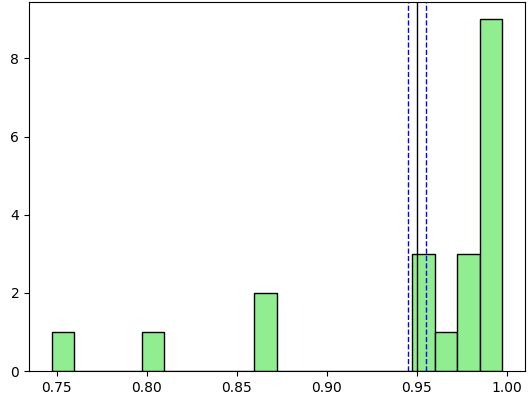}
    \caption{SSIM}
\end{subfigure}
\caption{Sample-wise performance of the naive NBRDF reconstruction with mean (solid) and variance (dashed line).}
\label{fig:hist_baseline}
\end{figure}

\begin{figure}
\centering
\begin{subfigure}{0.3\linewidth}
        \includegraphics[width=\linewidth]{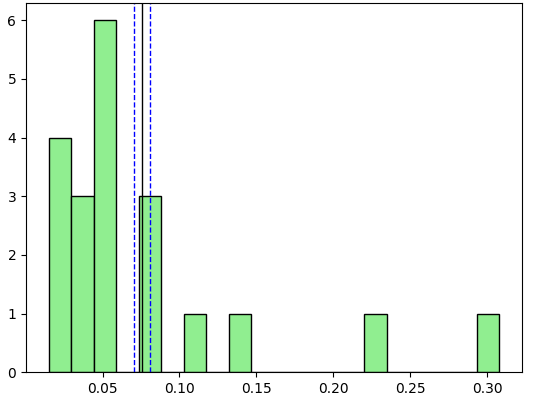}
    \caption{RMSE}
\end{subfigure}
\begin{subfigure}{0.3\linewidth}
        \includegraphics[width=\linewidth]{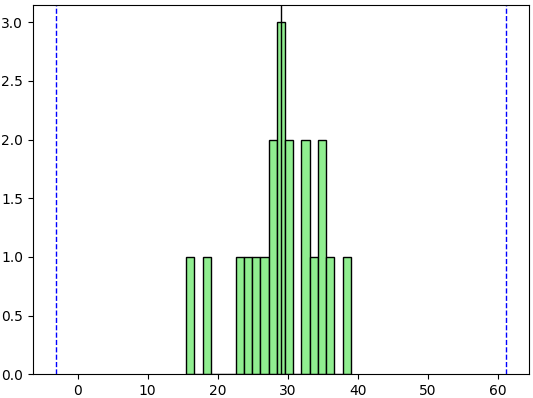}
    \caption{PSNR}
\end{subfigure}
\begin{subfigure}{0.3\linewidth}
        \includegraphics[width=\linewidth]{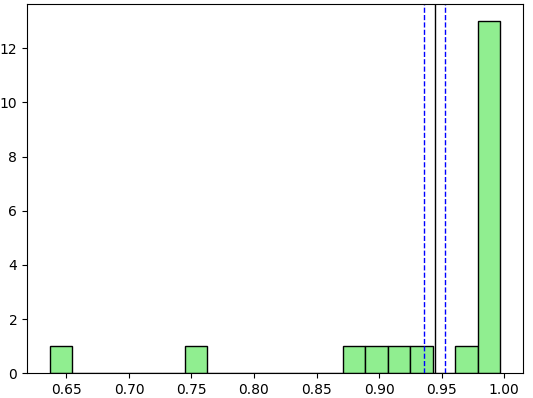}
    \caption{SSIM}
\end{subfigure}
\caption{Sample-wise performance of \ours on material reconstruction, with mean (solid) and variance (dashed line).}
\label{fig:hist_freq_kNN}
\end{figure}

We further visualize the sample-wise performance of the naive NBRDF~\cite{sztrajman2021nbrdf} reconstruction pipeline and \ours with frequency rectification in \cref{fig:hist_baseline,fig:hist_freq_kNN}, respectively, with the mean (solid line) and variance (dashed line) indicated. This visualization reveals overall statistical trends and helps identify outliers or failure cases, providing deeper insight into each model's behavior. The results show that \ours effectively reduces frequency domain discrepancies while maintaining comparable qualitative and quantitative performance. This underscores the potential of our method and suggests promising directions for future research in frequency-aware material modeling.

\section{Conclusion and Future Work} 
In this work, we propose \ours, a frequency-rectified neural material representation. Building on this representation, we develop a material reconstruction and editing pipeline based on an autoencoder architecture, incorporating a novel \emph{frequency-rectified loss} to enforce accuracy in the frequency domain. We evaluate our method against state-of-the-art baselines using both qualitative and quantitative metrics, with \ours consistently outperforming competing approaches. By integrating frequency information, \ours has the potential to unlock novel techniques for refining material details.

Promising directions for future work include extending the approach to more complex materials, such as anisotropic BRDFs, spatially varying BRDFs, and physically-based materials. We also hope this work contributes to future research and applications, including game development, virtual environments, and simulation systems, where realistic material representation is crucial for visual accuracy and immersion.

\small
\bibliographystyle{IEEEbib}
% \clearpage
\bibliography{main}

\normalsize
% \clearpage
% \appendix
% \section{Appendix}
% \input{sec/Appendix.tex}

\end{document}